\documentclass[preprint,prmaterials]{revtex4-1}
\usepackage {graphicx}
\usepackage{multirow}
\usepackage{amsmath}
\usepackage{siunitx}
\begin{document}

\raggedbottom

\title {Core-structure and lattice resistance of twinning dislocations in fcc metals}

\author {Sri Sadgun R. Pulagam and Amlan Dutta }

\affiliation {Department of Metallurgical and Materials Engineering, Indian Institute of Technology Kharagpur,
West Bengal 721302, India}

\date{\today}

\begin{abstract}
Metals with fcc structure may exhibit deformation twinning under specific conditions, which is an interesting but somewhat elusive aspect of their deformation behavior. It is well acknowledged that the phenomenon occurs through the activities of twinning partial dislocations. However, the lack of a comprehensive understanding of their fundamental properties obstructs the development of detailed multiscale models of crystal plasticity in the fcc metals. Here we explore the core-structures and lattice friction of twinning partials through atomistically informed numerical modeling. To this end, we choose four fcc crystals with widely differing stacking fault energies. Using the semi-discrete variational Peierls Nabarro model, we compute the core-widths and Peierls stresses of edge and screw twinning dislocations. Apart from the conventional layer-by-layer model of twin nucleation, the recently proposed alternate-shear model has also been examined. In the latter case, a negative stable fault energy has been observed, which is large enough to overcome the Peierls barrier. This study also highlights the significance of incorporating the surface correction, the absence of which leads to an overestimation of the intrinsic lattice resistance of the twinning dislocations.
\end{abstract}

\maketitle
\section{Introduction}
The competition between slip and twinning is a key phenomenon governing the plastic deformation of many metals and alloys with face-centered cubic (fcc) crystal structure. Conventionally, the general consensus has been that the deformation twinning is favored in metals like gold and silver, which have low stacking fault energies (SFEs) ~\cite{GoldSilver}. Also, several fcc alloys like Cu-Al alloys and TWIP steels exhibit twinning during deformation at room temperature ~\cite{CuAl, Twip}. Besides, even pure metals like Cu, Al, and Pd with moderate or high SFEs may undergo deformation twinning if they are in nanocrystalline forms~\cite{Cu-nanocrystal, Al-nanocrystal, Pd-nanocrystal}. Using both first-principles atomistic calculations and classical molecular dynamics simulations ~\cite{atomistic_calc,atomistic_calc2} it has been demonstrated that the propensity to deformation twinning in an fcc metal is dictated not only by the stable stacking fault energy but also by the unstable stacking and twin fault energies. It is well known that the deformation twinning initiates with the formation of a twin embryo, and the twin subsequently grows through the nucleation and growth of twinning dislocations or dislocation loops~\cite{GoldSilver, Twin_nucleation,Twin_nucleation2}. Although there are multiple possible mechanisms for the nucleation of twin embryo, the twinning dislocations always play a critical role in all of them ~\cite{reviewPaper}.\\

On account of the significance of twinning dislocations in the deformation mechanism maps of fcc metals and alloys, it is evident that their behavior and dynamics are of fundamental importance in the twinning process. Computational investigations of crystal plasticity often employ the dislocation mobility laws as key ingredients of multiscale modeling. In particular, a mesoscale method like discrete dislocation dynamics simulation inherently relies on the constitutive relations, which in turn, are extracted from the atomistic calculations~\cite{Mobility_dislocationDyn,Mobility_PureMetals, Mobility_Al, Mobility_bcc}. In this context, the core-structures of line defects become an indispensable contributor to such mobility relations by incorporating the effect of intrinsic lattice friction into the multiscale paradigm of deformation mechanism~\cite{cai2004dislocation, Orwan}. Due to its importance in various aspects of crystal plasticity, several approaches to the computation of core-structure and lattice friction have been developed. In recent years, a few studies have demonstrated the applicability of computational methods founded on the Frenkel-Kontorova~\cite{bcc_Finnis–Sinclairpotentials, Mod_of_edge_disl, Frenkel–Kontorovamodel, straight_mixed_disl} and phase-field~\cite{Phasefeild_fcc, Su_2019} models in exploring the core-structures. Nonetheless, most of the investigations conventionally employ computational models derived from the Peierls-Nabarro (PN) theory of dislocations. Over the last several years, new methods have been developed to enhance the reliability of atomistic calculations based upon the PN model. These include the semidiscrete-variational framework~\cite{SVPN_core, GSFE_Al}, development of non-local model~\cite{non-local_PN, non-local_PN2}, and incorporation of dislocation-phonon coupling~\cite{Wang_2008, comp_core}. Apart from the gradual improvements in the scheme of numerical implementation, these techniques have broadened the scope of studying the core properties in various materials, including metals, semiconductors, ceramics, minerals, etc. However, despite the well-recognized multiscale link between the dislocation core-structure and plastic deformation mechanism, very little information is available regarding the core-structures of twinning dislocations. Some preliminary calculations performed by Ogata \textit{et al.}~\cite{engy_landscape_bcc_fcc} followed a non-variational approach and one-dimensional generalized planar fault energies (GPFEs) to estimate the Peierls stress of twinning dislocations in Cu and Al. Nevertheless, a detailed and rigorous analysis remains unavailable. This becomes a hindrance against the development of accurate multiscale models of crystal plasticity, where the phenomenon of deformation twinning may play a critical role in the overall dynamics of deformation.\

Aiming to address the issue discussed above, we present a systematic investigation of the core-structure and lattice resistance of twinning dislocations in fcc metals. This study employs the semi-discrete variational Peierls Nabarro (SVPN) framework to compute the core-widths and Peierls stresses of edge and screw twinning dislocations in four metals with different stacking fault energies. In this regard, the glide of twinning dislocations have been classified into two categories - first, a twinning dislocation gliding over a well-formed twin boundary, thereby causing its migration, and second, the twinning dislocations associated with the nucleation of the initial twin embryos. In the latter category, we examine a newly proposed shear model of twin nucleation, in addition to the conventional layer-by-layer slip model. As the SVPN approach uses the GPFEs as inputs, they have been extracted from atomistic simulations using appropriate interatomic interaction models. \

\section{Atomistic simulation}
We construct virtual crystalline samples of Ag, Au, Cu, and Al with suitable embedded-atom-model (EAM) interatomic potentials ~\cite{EAM_Au_2005,EAM_Ag_Williams_2006,EAM_Cu_2001,EAM_Al_2008}, which have been fitted to the relevant material parameters obtained from the \textit{ab initio} calculations, and widely used for studying the mechanical properties through atomistic simulations. The two-dimensional generalized planar fault energy is computed by laterally sliding one part of the crystal over another part along the \{111\} atomic plane. The GPFE is measured as the variation in the sample's structural energy as a function of this two-dimensional disregistry. Figure 1 schematically depicts the three different slip configurations studied here. In the first (Fig. 1(a)), the initial structure consists of a fully developed twin slab, where the GPFE calculation involves a relative shift over a pre-existing twin boundary. A shift of $a\langle 11\bar{2} \rangle/6$ would cause the migration of the twin boundary by one atomic layer. We refer to this configuration as \textit{TG} (twin-growth) throughout the present text. The second case (Fig. 1(b)) corresponds to the layer-by-layer formation of a twin-nucleus, which has been the conventionally accepted twinning mechanism. Here the first shift is over a pre-existing intrinsic stacking fault (denoted as \textit{ABC}-1), which would create a single-layer meta-twin for a shift of $a\langle 11\bar{2} \rangle$/6. A subsequent shift over the consecutive plane (shown as \textit{ABC}-2 in the figure) converts the meta-twin into the two-layer twin embryo. The third case (Fig. 1(c)), which we refer to as the alternate-shear mechanism, has recently been demonstrated by Wang \textit{et al.}~\cite{ACB_case}. Here, the system consists of two pre-existing intrinsic stacking faults separated by one atomic layer. The shift is applied over the atomic layer between these two stacking faults, and the slip configuration is designated as \textit{ACB}. As shown in the figure, a twin embryo is thus formed for a disregistry of $a\langle 11\bar{2} \rangle$/6. \

\begin{figure}[t]
	\centerline{\includegraphics*[width=13cm, angle=0]{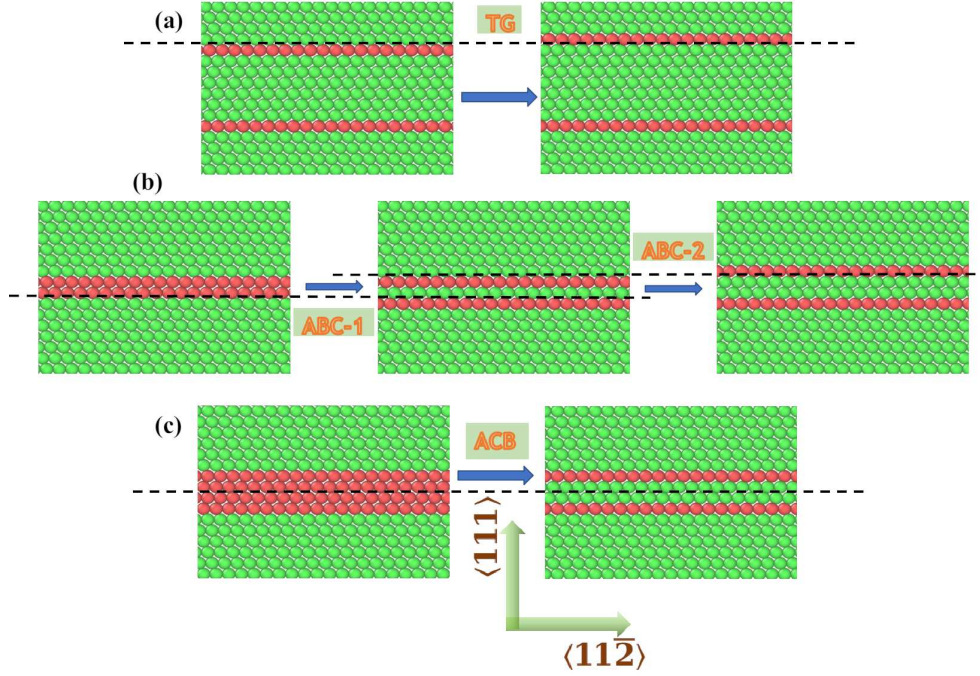}}
	\caption{(color online). Pictorial representation of the slip configurations, (a) twin-growth (\textit{TG}), (b) layer-by-layer (\textit{ABC}), and (c) alternate-layer (\textit{ACB}). The green particles indicate atoms with fcc structure, whereas the red ones have local hcp packing. A double-layer of hcp atoms represent an intrinsic stacking fault, whereas a single layer indicates a twin boundary. The dashed lines show the planes of disregistry, and the part of crystal above a disregistry plane glides over the part below it. }
\end{figure}

Periodic boundary condition is applied along the lateral directions, whereas the top and bottom free surfaces are created normal to the $\langle 111 \rangle$ direction. For every two-dimensional displacement, the system is relaxed to its minimum energy configuration, during which the atoms are allowed to move only normal to the plane of disregistry. The LAMMPS molecular dynamics code ~\cite{LAMMPS} has been employed to perform the molecular statics calculations, whereas the OVITO program ~\cite{OVITO} caters to visualization and other post-processing tasks.\

\section{Computation of core-structure and Peierls stress}
The variational Peierls-Nabarro framework offers a rigorous method of calculating the core-structure of a dislocation. It involves expressing the dislocation's energy as a functional of its disregistry field and tuning this field to minimize the energy functional. In particular, we employ the semi-discrete form of this approach ~\cite{SVPN_core}, which replaces the continuum integration by discrete summation over fixed lattice points, thereby rendering more accurate results. In this approach, the energy per unit length of a dislocation line is given by ~\cite{SVPN_core,SVPN_vs_Atomistics},
\begin{equation}
E_{disl} = E_{elastic} + E_{misfit} + E_{surface} + E_{stress}.       
\end{equation}
Here, $E_{elastic}$ refers to the elastic strain energy of the line defect and is computed as~\cite{GSFE_Al},	
\begin{equation}
E_{elastic} =  \sum_{i,j} \omega_{i,j}\left[K_{e}\rho_{i}^{e}\rho_{j}^{e}+K_{s}\rho_{i}^{s}\rho_{j}^{s}\right]  + Kb^{2}\ln R,
\end{equation}	
where $\omega_{i,j}=\frac{3}{2}\Omega_{i,i-1}\Omega_{j,j-1}+\Psi_{i-1,j-1}+\Psi_{i,j}-\Psi_{i,j-1}-\Psi_{j,i-1}$ ,with $\Omega_{i,j}=\textit{x}_{i}-\textit{x}_{j}$, and $\Psi_{i,j}=\frac{1}{2}\Omega_{i,j}^{2}ln|\Omega_{i,j}|$. The local misfit density of the $i^{th}$ nodal point at position $x_i$ is given by, $\pmb{\rho}_{i}=(\pmb u_{i}-\pmb u_{i-1})/(x_{i}-x_{i-1})$, with $\pmb u$ as the disregistry vector and $\rho_{i}^{e/s}$ as the corresponding edge/screw components.  $K_{s}=\mu/(4\pi)$ and $K_{e}=K_s/(1-\nu)$ are the pre-logarithmic energy factors for the edge and screw dislocations, respectively, where $\mu$ is the effective shear modulus, and $\nu$ is the Poisson's ratio. Within the domain of anisotropic elasticity theory, these effective elastic moduli have been computed for the \{111\} slip plane through the procedure outlined by Fitzgerald \textit{et al.} ~\cite{effctive_elst_const} with the stiffness constants obtained using the interatomic potentials. The mixed pre-logarithmic factor is given by,
\begin{equation}
K=\frac{\mu}{4\pi}\left(\frac{\sin ^{2}\beta}{1-\nu}+\cos ^{2}\beta\right),
\end{equation}
where $\beta$ is the characteristic angle of the dislocation. In Eq. (2), $R$ denotes the outer cut-off radius for the strain energy, although the second term on the LHS of the equation is irrelevant to the procedure of optimization.\

In Eq. (1), the misfit energy corresponding to disregistry across the slip plane is expressed as the following discrete summation~\cite{SVPN_core},
\begin{equation}
E_{misfit}=\Delta\textit{x}\sum_{i}\gamma\left(u_{i}^{e},u_{i}^{s}\right).
\end{equation}
Here, $\gamma(\pmb u)$ represents the 2-D misfit potential with $\textit{u}_{i}^{e/s}$ as the edge/screw components of the disregistry vector. Representing the disregistry vector as the superposition of two inverse-tangent terms is typically befitting in the case of a full dislocation, which dissociates into two Shockley partials ~\cite{Theory_of_disl}. In the present study, a single inverse-tangent term has been found to suffice as the twinning dislocation has the Burgers vector, $a\langle11\bar{2}\rangle/6$, identical to that of a single Shockley partial. The disregistry function satisfying  the constraints of the boundary conditions is expressed as,
\begin{equation}
u(x_{i})=\frac{b}{\pi}\arctan\frac{x_{i}}{\xi}+\frac{b}{2},
\end{equation}
where $b = a/\sqrt{6}$ is the Burgers vectors and $2\xi$ is the core-width of the dislocation. Also, $\textit{u}_{i}^{e}=\textit{u}(\textit{x}_{i})$ and $\textit{u}_{i}^{s}=0$ for an edge partial, and vice versa for the screw type.\

In the unified dislocation equation developed by Wang~\cite{Wang_2008}, it has been demonstrated that the effect of discreteness of the lattice can be incorporated in the original PN model by adding a second-order differential term. This introduces the so-called surface correction given by~\cite{SVPN_vs_Atomistics},
\begin{equation}
E_{surface}= \frac{1}{4}\sum_{i}\left(B_{e}\left(\rho^{e}_{i}\right)^{2}+B_{s}\left(\rho^{s}_{i}\right)^{2}\right) \Delta x
\end{equation}
where, $B_{e}$ and $B_{s}$ are calculated as,
\begin{equation}
B_{e}=\frac{3d}{4}\left(c_{l}-c_{tv}\tan^{2}\theta\cos^{2}\phi\right)
\end{equation}
\begin{equation}
B_{s}=\frac{3d}{4}\left(c_{th}-c_{tv}\tan^{2}\theta\sin^{2}\phi\right)
\end{equation}
In the above relationships, $d = a/\sqrt{3}$, is the interplanar spacing between adjacent \{111\} planes. $\theta$ and  $\phi$ are the orientation angles associated with the relative pair of neighbor atom vectors and depend on the crystal structure. In the case of fcc crystals with the \{111\} slip plane, $\tan\theta=1/\sqrt{2}$ and $\phi=\pi/6$. For phonons propagating in the $\langle 1 1 0\rangle$ direction, the effective elastic constant, which is related to the longitudinal wave velocity is obtained as~\cite{SVPN_vs_Atomistics},
\begin{equation}
c_{l}=\frac{C_{11}+C_{12}+2C_{44}}{2}.
\end{equation}
Similarly, the effective moduli $c_{th}$ and $c_{tv}$ correspond to the resolved polarized components of the transverse waves, and are estimated as,
\begin{equation}
c_{th}=\frac{C_{11}-C_{12}}{2}, 
\end{equation}
\begin{equation}
c_{tv}=C_{44}.
\end{equation}
The effective shear modulus ($\mu$), Poisson's ratio ($\nu$), and the elastic constants $c_{l}$, $c_{th}$, and $c_{tv}$ of all the four metals are computed and presented in Table I. \
\begin{table}[t]
	\caption{Effective stiffness constants of the four fcc metals.}
		\begin{tabular}{|c|c|c|c|c|}
			\hline
			& \textbf{Ag} & \textbf{Au} & \textbf{Cu} & \textbf{Al} \\
			\hline
			$\boldsymbol{\mu }$(GPa) & 27.8 & 28.4 & 44.6 & 26.5\\
			\hline
			$\boldsymbol{\nu}$ & 0.43 & 0.48 & 0.41 & 0.36\\
			\hline
			$\mathbf{c_{l}}$(GPa) & 155.5 & 231.6 & 222.4 & 112.9\\
			\hline
			$\mathbf{c_{th}}$(GPa) & 15.2 & 16.1 & 23.6 & 22.8\\
			\hline
			$\mathbf{c_{tv}}$(GPa) &46.4 &45.9 &76.2 &30.7\\
			\hline
		\end{tabular}
\end{table}

Finally, the change in enthalpy of the system on account of applied shear stress is given by~\cite{non-local_PN2},
\begin{equation}
E_{stress}=-\frac{1}{2}\sum_{i}\left[\{u^{e}(x_{i})+u^{e}(x_{i+1})\}\tau^{e}+\{u^{s}(x_{i})+u^{s}(x_{i+1})\}\tau^{s}\right]\Delta x,
\end{equation}

\noindent where $\tau^{e}$  and $\tau^{s}$ are the shear stresses interacting with the disregistry components, $u^{e}(x)$ and $u^{s}(x)$, respectively.  \

The total energy functional is minimized by initially tuning the core-structure using a global Bayesian optimizer ~\cite{Bayesian} with the Mat\'{e}rn-5/2 covariance kernel ~\cite{Gaussian}. This is followed by a cascade of local optimizations comprising quasi-newton, simplex-search, and conjugate-gradient methods. As the numerical solution of the variational PN model can be sensitive to the choice of initial conditions, this rigorous optimization protocol is aimed at enhancing the reliability and reproducibility of the optimal parameters. For evaluating the Peierls stress, the optimization is carried out at gradually increasing values of the applied shear load, thereby altering the $E_{stress}$ as computed in Eq. (12). The critical stress at which the optimizers fail to converge is noted as the Peierls stress of the twinning dislocation ~\cite{GSFE_Al}.\

\section{Results and discussion}

\subsection{Layer-by-layer and twin-growth mechanisms}
\begin{figure}[t]
	\centerline{\includegraphics*[width=8cm, angle=0]{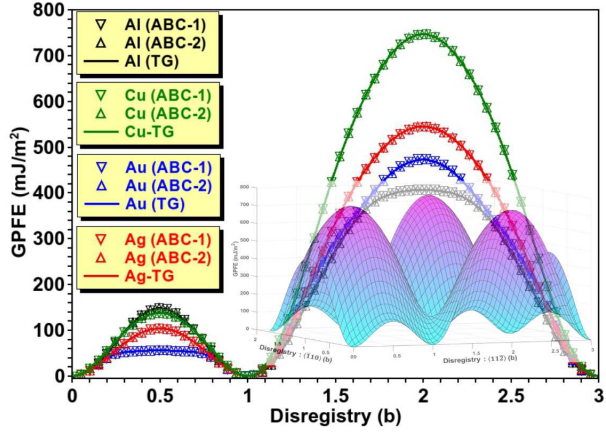}}
	\caption{(color online). The GPFE profiles for disregistry along $\textbf{b} = a<11\bar{2}>/6$ for the layer-by-layer (\textit{ABC}-1 and \textit{ABC}-2) and twin-growth (\textit{TG}) slip configurations. The first and second local maxima represent the first and second unstable twin fault energies, respectively. A typical two-dimensional GPFE profile is also demonstrated as the transluscent surface plot superimposed on the one-dimensional profile.}
\end{figure}
Although the PN model's numerical implementation employs the two-dimensional GPFE, it is easier to visualize its features through its one-dimensional projection. The one-dimensional generalized planar fault energies for the disregistry along the $\langle11\bar{2}\rangle$ direction are plotted in Fig. 2, which displays the variation in energies for misfits corresponding to the slip configurations \textit{TG}, \textit{ABC}-1, and \textit{ABC}-2 as depicted in Fig. 1. The first remarkable observation is that for each of the four metals, the corresponding plots for the three slip configurations coincide. This can be attributed to the fact that the unstable and stable planar fault energies for a closed-packed atomic plane are dominated by the first and second nearest-neighbor interatomic interactions. For the \{111\} plane of fcc, the stacking sequence across the boundary of a full twin (configuration \textit{TG} in Fig. 1) is of type \textit{BC}\textbar \textit{BA}. Accordingly, the second nearest-neighbor interactions dominating the fault energy are (\textit{B-B}) and (\textit{C-A}). A close inspection of the stacking sequences reveals that crystallographically equivalent interactions are present in the slip configurations \textit{ABC}-1 and \textit{ABC}-2 as well, which explains the identical GPFE profiles for all the three slip configurations as seen in Fig. 2. As the planar fault energies of slip configurations \textit{ABC}-1, \textit{ABC}-2, and \textit{TG} are found to coincide, we shall consider them together for the subsequent computations of dislocation properties.\

Another noteworthy feature of the GPFE is its qualitative distinction from a typical GSFE profile. Akin to the generalized stacking fault energy, the GPFE shows two energy maxima corresponding to the unstable fault energies. However, unlike the GSFE, which exhibits the characteristic stable stacking fault energy at a disregistry of $a\left\langle 11\bar{2}\right\rangle/6$, the GPFE profiles in Fig. 2 indicate the invariance of fault energy for the same disregistry. This is because the disregistry in the former case entails the formation of a stacking fault in an otherwise perfect crystal, whereas the latter case merely involves the migration of a pre-existing twin boundary from one slip plane to the adjacent one. We also note that the first unstable twin fault energy follows the order, Au $<$ Ag $<$ Cu $\lesssim$ Al, whereas the second one follows, Al $<$ Au $<$ Ag $<$ Cu.\

\begin{figure}[t]
	\centerline{\includegraphics*[width=15cm, angle=0]{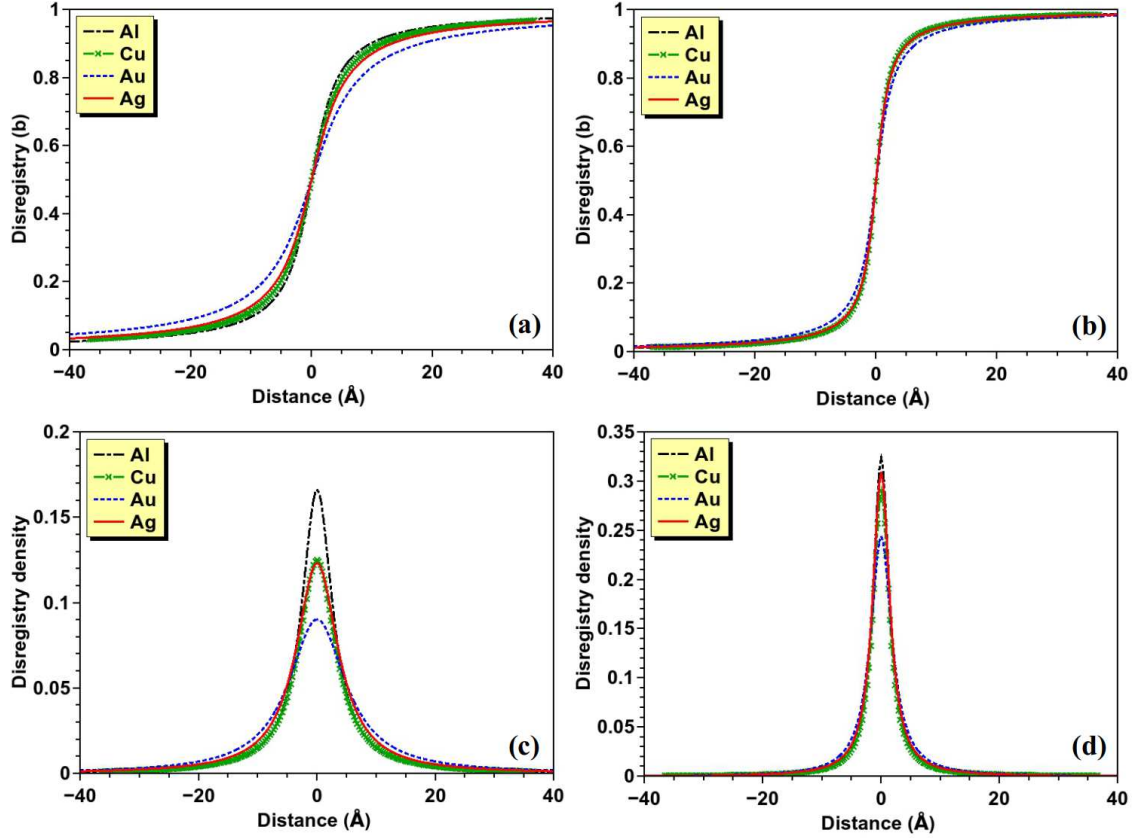}}
	\caption{(color online). Disregistry profiles of (a) edge and (b) screw twinning partials involved in the layer-by-layer formation and twin-growth mechanisms. The corresponding disregistry densities are also shown for the (c) edge and (d) screw type line defects. }
\end{figure}

Figures 3(a) and (b) show the disregistry distributions of the edge and screw defects, respectively, while the corresponding local dislocation densities are plotted in Figs. 3(c) and (d). The screw-type twinning partials are observed to have narrower cores as compared to their edge-type counterparts, thereby suggesting larger Peierls stresses. The results are summarized in Table-II, which displays the core-widths (2$\xi$) and Peierls stresses in the four fcc metals. For both edge and screw partials, Au exhibits the widest cores and smallest Peierls stresses, whereas Al has the narrowest cores and the largest Peierls stress.\

\begin{table}[ht]
	\caption{Computed core-widths and Peierls stresses of twinning dislocations associated with the \textit{ABC}-1, \textit{ABC}-2, and \textit{TG} slip configurations. For comparison, the values estimated without including the surface corrections are also shown with the asterisks (*).}
	\scalebox{1}{
	\begin{tabular}{|c|c|c|c|c|c|c|c|c|}
	\cline{2-9}
	\multicolumn{1}{c|}{}& \multicolumn{2}{c|}{\textbf{Ag}} & \multicolumn{2}{c|}{\textbf{Au}} & \multicolumn{2}{c|}{\textbf{Cu}} & \multicolumn{2}{c|}{\textbf{Al}} \\ \cline{2-9}
	\multicolumn{1}{c|}{}&2$\boldsymbol{\xi}$  & $\boldsymbol{\tau_{p}}$  & 2$\boldsymbol{\xi}$  & $\boldsymbol{\tau_{p}}$  & 2$\boldsymbol{\xi}$  & $\boldsymbol{\tau_{p}}$  & 2$\boldsymbol{\xi}$  & $\boldsymbol{\tau_{p}}$  \\ 
	\multicolumn{1}{c|}{}& (\AA) &  (MPa) &  (\AA) &  (MPa) & (\AA) &  (MPa) & (\AA) &  (MPa) \\ \cline{1-9}
	EDGE& 8.6/$4.8^{\ast}$ & 10/$19^{\ast}$ & 11.7/$7.4^{\ast}$ & 4/$8^{\ast}$ & 7.5/$4.8^{\ast}$ & 20/$34^{\ast}$ & 6.3/$4.0^{\ast}$ & 29/$40^{\ast}$\\
	\hline
	SCREW & 3.4/$3.0^{\ast}$ & 47/$70^{\ast}$ & 4.3/$3.6^{\ast}$ & 19/$23^{\ast}$ & 3.2/$2.8^{\ast}$ & 53/$89^{\ast}$ & 3.2/$2.3^{\ast}$ & 58/$107^{\ast}$\\
	\hline
	\end{tabular}}
\end{table}

\subsection{Alternate layer mechanism}
Having analyzed the generalized planar fault energies and core-structures of twinning partials for the layer-by-layer and twin-growth slip configurations, we now consider the slip corresponding to the \textit{ACB} shear depicted in Fig. 1(c). Unlike the previous cases, where the slip of $a\left\langle 11\bar{2}\right\rangle/6$ merely causes the twin-boundary to migrate, a similar slip in the \textit{ACB} model creates a different stacking altogether. This is suggestive of a stable fault energy along the lines of the stacking fault energy obtained in a typical GSFE plot. We also note that the mechanism of \textit{ACB} shear has two stacking faults on alternate slip planes, which effectively assumes the structure of a four-layer thin slab of hcp atoms. As the relative slip between these two stacking faults entails reverting of the two middle atomic layers back to the stable fcc structure, it is clearly expected to reduce the system's structural energy. As a result, the change in stacking sequence caused by the slip, as seen in Fig. 1(c), leads us to predict a negative value of the stable fault energy.\

\begin{figure}[t]
	\centerline{\includegraphics*[width=9cm, angle=0]{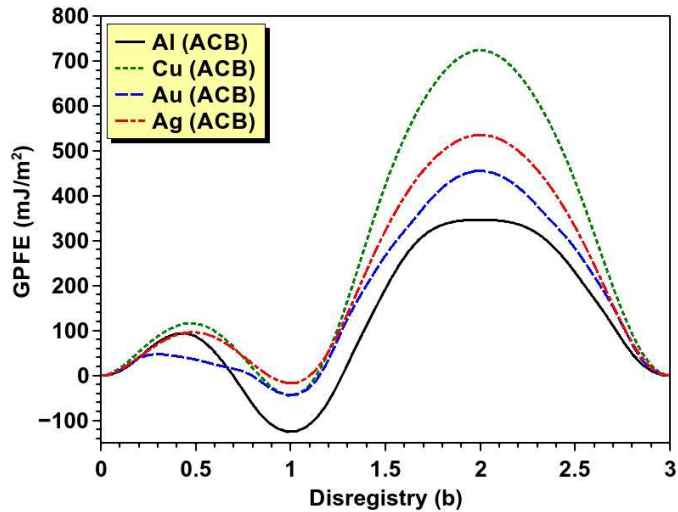}}
	\caption{(color online). One-dimensional GPFE for disregistry along $\textbf{b}=a<11\bar{2}>/6$ for the alternate-layer (\textit{ACB}) slip configurations. }
\end{figure}

Figure 4 displays the GPFE profiles for the slips in the \textit{ACB} model. We immediately notice that for all the four metals, they indeed reveal negative fault energies for a disregistry of $a<11\bar{2}>/6$, albeit to different extents. The magnitude of this fault energy follows the order Al $\gg$ Cu $\approx$ Au $>$ Ag, which as expected, correlates with the differences in cohesive energies of the hcp and fcc phases of these metals (Al: 27.6 meV, Cu: 7.8 meV, Au: 9.5 meV, Ag: 3.8 meV). On account of a fault energy of significant magnitude, computation of the core-structure and Peierls stress of the partial becomes non-trivial. Here the negative fault energy favors the expansion of the fault area, and thereby exerts a large forward stress on the twinning partial. As a consequence, the dislocation moves on its own by overcoming the Peierls barrier. One may argue that in such a case, the notion of Peierls stress becomes irrelevant in its conventional sense, for an isolated partial would invariably be associated with a fault and appear to move on its own without requiring the assistance of an externally applied load. Another way of perceiving this effect is by considering the fact that the fault energy is already a part of $E_{misfit}$ (Eq. (4)), which in turn is an inseparable part of the overall energetics and Peierls barrier by virtue of Eq. (1). However, we also realize that despite the presence of this fault stress, the lattice friction is always present due to the intrinsic discreteness of the lattice. Hence, the intrinsic stress on the partial dislocation exerted by the fault acts against this lattice resistance while pushing the dislocation in the forward direction. In principle, this resistance should be quantifiable in terms of an effective Peirls stress, for the actual Peierls stress is essentially zero, and should be separable from the net stress exerted on the partial.\

As the stress exerted by the fault far exceeds the intrinsic lattice friction and causes the partial to move, minimizing the energy of the system (Eq. 1) leads to failure of convergence even without any external shear stress. Here we devise a simple computational strategy, which allows the calculation of the core-width and the effective Peierls stress on the isolated partial dislocation in an indirect manner. In the absence of an applied load, the stacking fault with negative fault energy applies a fault stress, $\tau_{f}$. This stress pushes the dislocation in the forward direction, while the effective Peierls stress, $\tau^{\prime}_{p}$, acts in the opposite direction (\textit{c.f.} Fig. 5(a)). If $\tau_{f} > \tau^{\prime}_{p}$, the dislocation becomes unstable and keeps moving in the forward direction, as seen in our case of the \textit{ACB} shear. Therefore, in the present technique, we add an external stress, $\tau_1$ in the direction opposite to $\tau_{f}$, and increase it till the energy minimizer converges and the dislocation becomes stable. The stress balance on the dislocation line yields (Fig. 5(a)),
\begin{figure}[t]
	\centerline{\includegraphics*[width=6.5cm, angle=0]{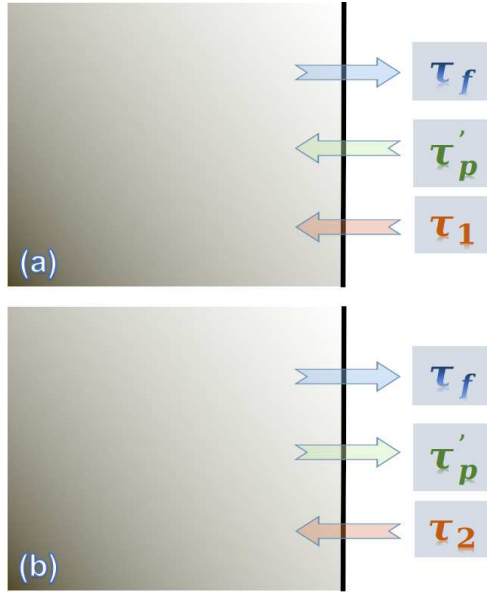}}
	\caption{(color online). Unit cube of the model material showing spherical pinning defects and dislocation lines. }
\end{figure}

\begin{equation}
\tau_1 = \tau_{f} - \tau^{\prime}_{p}.
\end{equation}

\begin{table}[b]
	\centering
	\caption{Core-widths ($2\xi$), effective Peierls stresses ($\tau^{\prime}_{p}$), and fault stresses ($\tau_f$) in the \textit{ACB} slip configuration. The last column presents the fault stresses ($\tau_f^{\Gamma}$) computed as $\Gamma/b$. The values with astersks (*) correspond to the results obtained without accounting for the surface correction. }
	\scalebox{1.1}{
		\begin{tabular}{|c|c|c|c|c|c|c|c|} \cline{2-8}
			\multicolumn{1}{c|}{}& \multicolumn{2}{c|}{2$\boldsymbol{\xi}$}  & \multicolumn{2}{c|}{$\boldsymbol{\tau_{f}}$} & \multicolumn{2}{c|}{$\boldsymbol{\tau^{\prime}_{p}}$} & $\boldsymbol{\tau_{f}^{\Gamma}}$ \\ 
			\multicolumn{1}{c|}{}& \multicolumn{2}{c|} {(\AA)}  & \multicolumn{2}{c|}{(MPa)} & \multicolumn{2}{c|}{(MPa)} & (MPa) \\ \cline{2-7}
			\multicolumn{1}{c|}{} & EDGE & SCREW & EDGE & SCREW &EDGE & SCREW & \\ 
			\hline 
			\textbf{Ag}&8.7/$5.4^{\ast}$ & 3.4/$2.8^{\ast}$ & 93/$107^{\ast}$ & 117/$94^{\ast}$ & 19/$49^{\ast}$ & 55/$74^{\ast}$ & 98.4 \\
			\hline
			\textbf{Au} & 11.8/$7.2^{\ast}$ & 4.2/$3.6^{\ast}$ & 259/$250^{\ast}$ & 272/$250^{\ast}$ & 14/$25^{\ast}$ & 49/$40^{\ast}$ & 255.8 \\
			\hline
			\textbf{Cu} & 7.5/$4.8^{\ast}$  & 3.2/$2.7^{\ast}$  & 293/$298^{\ast}$  & 321/$288^{\ast}$& 30/$60^{\ast}$ &71/$111^{\ast}$ & 296.3 \\
			\hline
			\textbf{Al} & 6.4/$3.8^{\ast}$  & 3.3/$2.3^{\ast}$  & 743/$757^{\ast}$  & 773/$789^{\ast}$ & 22/$54^{\ast}$ & 138/$187^{\ast}$ & 754.1 \\
			\hline
	\end{tabular}}
	
\end{table}

Once $\tau_1$ is noted, the external stress is further increased until it attains a critical value $\tau_2$, at which the partial dislocation moves in the backward direction and the fault region starts shrinking. This situation is easily identified, as the energy optimizer again fails to converge if the applied shear load exceeds $\tau_2$. Clearly, $\tau_2$ is now acting against both the lattice friction and the fault stress (Fig. 5(b)), and we have,
\begin{equation}
\tau_2 = \tau_{f} + \tau^{\prime}_{p}.
\end{equation}
\indent From Eqs. (13) and (14), the fault stress is calculated as, $\tau_f = \left(\tau_1 + \tau_2\right)/2$, while the effective Peierls stress is obtained as, $\tau^{\prime}_{p} = \left(\tau_2 - \tau_1\right)/2$. Table-III presents the values of  $\tau_f$ and $\tau^{\prime}_{p}$, along with the core-widths computed using this technique. As the $\tau_f$ values are slightly larger for the screw partials than those for the edge type, the table displays the estimated $\tau_f$ as the mean of the corresponding screw and edge values. To test the fundamental idea of separating an effective Peierls stress and the validity of the aforementioned method to do so, we note that the stress exerted by the fault on the twinning partial in the \textit{ACB} shear mechanism is roughly given by, $\tau_f^{\Gamma} = \Gamma/b$, where $\Gamma$ is the magnitude of the negative fault energy, and $b = a/\sqrt{6}$ is the length of the Burgers vector of the twinning partial. Table-III also shows the $\tau_f^{\Gamma}$ computed from the fault energy, and a direct comparison with $\tau_f$ obtained by solving Eqs. (13) and (14) reveal them to be sufficiently close to each other, thus justifying the resolution of the stresses into various components, as shown in Fig. (5).\

\subsection{Effect of surface correction}
The original PN model did not inherently account for the lattice discreteness, and the variational approach can, in principle, be implemented within a continuum framework, albeit with severe inaccuracy in the computational results. The semi-discrete variational method was the first rigorous step in introducing the effect of lattice discreteness by replacing the integration over the spatial domain with summation over lattice sites. However, the surface correction (Eqs. (6-11)), which depends on the acoustic phonon velocity and crystal structure of the material, incorporates an additional effect of discreteness with a robust physical and analytical approach. As this correction is newly developed and still not included in many subsequent studies involving the PN modeling, it is pertinent to estimate its significance in determining the core-structure and Peierls stress of the twinning dislocations. To this end, we remove the surface correction term (Eq. 6) from the total energy of dislocation, and recompute the results for both \textit{ABC} and \textit{ACB} shear models. The core-widths and Peierls stresses obtained without the surface correction are given in Tables-II and III (values with the asterisks), alongside the values computed from the full model. A direct comparison shows that the core-width is underestimated in the absence of the surface correction, while the Peierls stress in overestimated. Moreover, for most of the systems, the Peierls stress is overestimated by a large extent, suggesting that the quantitative modeling of fcc twinning partials is not feasible by ignoring the surface correction.\

\section{Conclusion}
We employ atomistically informed Peierls-Nabarro modeling to examine the core-structures and Peierls stresses of twinning partial dislocations in fcc metals. This study includes twinning partials involved in two different mechanisms of twin formation. The semi-discrete variational framework adopted here relies on the generalized fault energy estimated from the atomistic computations. We find that the GPFEs considered here exhibit qualitative dissimilarity as compared to the typical GSFE profile of the $\{111\}$ plane in fcc metals. The slip associated with the layer-by-layer twinning mechanism merely causes twin-boundary migration without forming a new planar fault, whereas the slip in the alternate-layer mechanism produces a new fault with negative stable fault energy. In the latter case, estimation of the lattice friction proves to be non-trivial, and we demonstrate a new strategy to obtain the effective Peierls stress. Our calculations also underline the importance of incorporating the surface correction and reveal that its omission may lead to severe inaccuracy in estimating the core-width and lattice friction of a twinning partial.\

Despite the well-known importance of twinning dislocations in dictating the deformation mechanism maps, their fundamental properties have largely remained unexplored. The qualitative difference between the twinning GPFE and the GSFE, particularly in the alternate-shear mechanism, suggests that the other dislocation properties depending on the stable and unstable fault energies, also merit close examination. For instance, the modern computational models suggest that akin to the core-structure, the energy barrier and critical stress for the nucleation of twinning dislocation loops should also depend on the generalized planar fault energy ~\cite{conc_1,conc_2}. Similarly, the twinning screw partial in metals with bcc crystal structures can be a system of interest on account of the peculiar behavior of screw dislocations in bcc metals ~\cite{conc_3,conc_4}. Thus it is reasonable to expect that the results obtained here would motivate further studies to fit the atomistic information in the larger picture of multiscale modeling.

\section*{Acknowledgements}
AD is thankful to the Department of Science and Technology (Government of India), and the Indian Institute of Technology Kharagpur, for providing financial support to this work under the ECRA and ISIRD grants, respectively. SR thanks the Tata Consultancy Services for providing financial assistance through the TCS scholarship program.

%\bibliography{ref}

\end{document}